\documentclass[a4paper,12pt]{article}

\usepackage{amsmath}
\usepackage{amssymb}
\usepackage{fullpage}
\usepackage{verbatim}

\usepackage[T1,OT1]{fontenc}

\usepackage{graphicx}

\allowdisplaybreaks[3]

\usepackage[hang]{footmisc}
\usepackage[compress,square,numbers]{natbib}
\usepackage[colorlinks,linktocpage,linkcolor=blue,citecolor=blue,urlcolor=blue]{hyperref}

\renewcommand{\d}{\mathrm{d}}
\newcommand{\e}{\mathrm{e}}

\newcommand{\nl}{\notag \\ &\quad\,}

\begin{document}

\numberwithin{equation}{section}

\thispagestyle{empty}

\begin{flushright}
\normalsize
\end{flushright}
\vspace*{1cm}

\begin{center}

{\LARGE \bf On SUSY-breaking Moduli Spaces}

\vspace{0.5cm}

{\LARGE \bf of AdS$_7$ Vacua and 6D SCFTs}

\vspace{2cm}
{\large Daniel Junghans${}^{1}$ and Marco Zagermann${}^{2}$}\\

\vspace{1cm}
${}^1$ Institut f{\"{u}}r Theoretische Physik, Ruprecht-Karls-Universit{\"{a}}t Heidelberg,\\
Philosophenweg 19, 69120 Heidelberg, Germany\\

\vspace{0.5cm}
${}^2$ Fachbereich Physik der Universit\"{a}t Hamburg,\\
Luruper Chaussee 149, 22761 Hamburg, Germany\\

\vspace{1cm}
{\upshape\ttfamily junghans@thphys.uni-heidelberg.de, marco.zagermann@desy.de}\\

\vspace{1.5cm}

\begin{abstract}
\noindent We study supersymmetric AdS$_7$ vacua of massive type IIA string theory, which were argued to describe the near-horizon limit of NS5/D6/D8-brane intersections and to be holographically dual to 6D $(1,0)$ theories. We show, for the case without D8-brane charges, that such vacua do not admit any supersymmetry-breaking deformations. Our result implies that the dual $(1,0)$ theories do not have a conformal manifold, thus extending previously known results for supersymmetric deformations. It is also in line with the recent conjecture that non-supersymmetric AdS vacua are in the swampland.
\end{abstract}

\end{center}

\newpage

\section{Introduction}

AdS vacua in string theory are of both formal and phenomenological interest. They are among the simplest vacua one can construct in string theory and thus serve as toy models for moduli stabilization. Indeed, more realistic solutions with positive vacuum energy are often argued to exist based on uplifting AdS vacua \cite{Kachru:2003aw, Balasubramanian:2005zx}. Furthermore, AdS vacua are extensively studied in the context of the gauge-gravity correspondence \cite{Maldacena:1997re, Gubser:1998bc, Witten:1998qj}.

In this note, we study AdS$_7$ vacua of massive type IIA string theory. These vacua are obtained by compactifying on a space with the topology of a 3-sphere and supported by $H_3$-flux and D6/D8-branes. The simplest setup with only D6-branes was first studied in \cite{Blaback:2011nz, Blaback:2011pn} (see also \cite{Blaback:2010sj}), where the non-supersymmetric smeared solution\footnote{The breaking of supersymmetry in this solution was discussed in \cite{Danielsson:2013qfa}.} was constructed and some properties of solutions with localized brane distributions were discovered. It was later realized in \cite{Apruzzi:2013yva} that the setup with localized branes also admits supersymmetric vacua, which were constructed numerically. In a second version of \cite{Apruzzi:2013yva}, also supersymmetric solutions with D8-branes were presented. Such solutions can alternatively be understood as a consequence of brane polarization \cite{Junghans:2014wda}. Finally, in \cite{Apruzzi:2015zna}, the general supersymmetric solution was found analytically.
It was argued to be holographically dual to 6D $(1,0)$ theories \cite{Gaiotto:2014lca, Cremonesi:2015bld, Apruzzi:2017nck, DeLuca:2018zbi}, which arise from intersecting NS5/D6/D8-brane systems \cite{Intriligator:1997kq, Intriligator:1997dh, Brunner:1997gf, Hanany:1997gh, Janssen:1999sa, Imamura:2001cr, Bobev:2016phc, Macpherson:2016xwk}.
Consistent truncations of massive type IIA around supersymmetric AdS$_7$ vacua were studied in \cite{Passias:2015gya, Malek:2018zcz, Malek:2019ucd}.

Contrary to the supersymmetric case, non-supersymmetric solutions in this setup have not been fully classified. In addition to the smeared solution \cite{Blaback:2011nz}, it is known that non-supersymmetric solutions with localized brane profiles exist \cite{Dibitetto:2015bia, Passias:2015gya, Apruzzi:2016rny}, but it is not clear whether they are the only ones. In \cite{Junghans:2014wda}, a numerical scan for vacua of the setup was conducted. The results of the scan hinted at the possibility of a family of non-supersymmetric solutions that are continuously connected to the supersymmetric one via a flat direction in the scalar potential.\footnote{In 4D Minkowski string vacua, supersymmetry cannot be continuously broken \cite{Dine:1987vf, Banks:1988yz}, but we are not aware of arguments forbidding this in a 7D AdS context.} However, the evidence was not conclusive since the search was carried out in a multi-dimensional parameter space with singular boundary conditions. This made it difficult to control numerical errors, in particular concerning the precise brane content of the deformed solutions. In this note, we therefore study deformations of the supersymmetric solutions analytically.

A common lore is that non-supersymmetric moduli spaces are unlikely since, in the absence of supersymmetry, nothing protects the moduli from being lifted. However, flat directions in moduli space often emerge in certain limits (for example, classically) even if they are not exact. Furthermore, it was argued in \cite{Bashmakov:2017rko} that non-supersymmetric CFTs with a gravity dual can admit conformal manifolds beyond the planar limit. This would correspond to a moduli space for non-supersymmetric AdS vacua that survives the inclusion of (at least some) quantum corrections. Here, we analyze supersymmetry-breaking deformations of supersymmetric AdS$_7$ vacua in the classical type IIA supergravity limit. The main result of this paper is that such deformations do not exist. Our result thus shows that, already in the classical limit, the vacua do not have any moduli space.
Via the holographic correspondence, this implies
that the dual 6D $(1,0)$ theories do not have a conformal manifold.\footnote{Strictly speaking, this argument does not rule out non-supersymmetric marginal deformations of the $(1,0)$ CFT that do not have a holographic dual supergravity description (cf.\ also the discussion in \cite{Ooguri:2016pdq}).} Previous studies of AdS$_7$ moduli spaces and the conformal manifolds of the dual CFTs were concerned with supersymmetry-preserving deformations and found that the AdS vacua are isolated \cite{Louis:2015mka, Cordova:2016xhm}. Our result extends these analyses to non-supersymmetric deformations.

Our result is also in line with the recent conjecture that non-supersymmetric AdS solutions must be unstable \cite{Ooguri:2016pdq, Freivogel:2016qwc, Danielsson:2016mtx}. If there was a family of non-supersymmetric AdS solutions continuously connected to the supersymmetric one, perturbatively stable non-supersymmetric AdS vacua would exist infinitesimally close to the supersymmetric one unless the latter has a tachyon exactly at the Breitenlohner-Freedmann bound. Since these non-supersymmetric vacua would have the same vacuum energy as the supersymmetric one, they would also have a good chance of being non-perturbatively stable and thus be promising candidate examples to violate the AdS conjecture. However, our result shows that no such deformations of the supersymmetric vacua exist, thus lending further support to the conjecture in this specific setup.

This paper is organized as follows. In section \ref{sec:setup}, we review the setup and establish our conventions. In section \ref{sec:def}, we consider linearized deformations around a supersymmetric solution and derive an ODE system determining them. In sections \ref{sec:ode} and \ref{sec:ode2}, we solve the ODE system and show that it only admits the trivial solution where all deformations are zero. In section \ref{sec:concl}, we summarize our results and discuss possible directions for future research.

\section{Setup}
\label{sec:setup}

We consider AdS$_7$ solutions of massive type IIA string theory. The general ansatz for the Einstein frame metric is\footnote{The solutions studied in this paper have pointlike sources at one or both poles of the $S^3$. The backreaction of these sources preserves an SO(3) isometry, i.e., the field profiles are invariant under rotations around the poles \cite{Blaback:2011nz}.} \cite{Blaback:2011nz, Blaback:2011pn}
\begin{equation}
\d s_{10}^2 = \e^{2A(y)}\d s_7^2 + \e^{2B(y)} \ell_\text{s}^2 \left( \d y^2 + \e^{2C(y)}\d s_2^2 \right),
\end{equation}
and the form fields can be parametrized as
\begin{equation}
H_3 = \lambda(y) F_0 \e^{7/4\phi(y)}\star_3 1, \quad F_2 = \e^{-3/2\phi(y)-7A(y)} \star_3 \d \left(\lambda(y)\e^{3/4\phi(y)+7A(y)}\right).
\end{equation}
Here, $\d s_7^2$ is the unwarped AdS$_7$ metric with radius $\ell_\text{s}=\sqrt{\alpha^\prime}$, $\d s_2^2$ is the metric of a round unit 2-sphere and $y$ is a (dimensionless) angle. The function $C(y)$ may be fixed by a suitable coordinate transformation, but it is convenient to keep it arbitrary for now. For later convenience, we also introduce the flux integers
\begin{equation}
f_0 = 2\pi \ell_\text{s} F_0 \in \mathbb{Z}, \qquad h=\frac{1}{(2\pi)^2\ell_\text{s}^2}\int_{S^3}H_3 \in \mathbb{Z}. \label{flux}
\end{equation}
From now on, we will assume $f_0>0$ without loss of generality.

The class of supersymmetric AdS$_7$ vacua constructed in \cite{Apruzzi:2015zna} can be written in terms of a single function $\beta(y)$ as follows:
\begin{align}
\e^{2A} &= \frac{8 \left(4\beta-y\beta^\prime\right)^{1/4}}{3\sqrt{3}} \left(\frac{-\beta^\prime}{y}\right)^{-1/8}, \label{susysol1} \\ \e^{2B} &= \frac{ \left(4\beta-y\beta^\prime\right)^{1/4}}{3\sqrt{12}\beta}\left(\frac{-\beta^\prime}{y}\right)^{7/8}, \\ \e^{2C} &= 4 \left(\frac{-\beta^\prime}{y}\right)^{-1} \frac{\beta^2}{4\beta-y\beta^\prime} = \left(\frac{2}{3}\right)^4 \e^{-2A-4B+1/2\phi}, \\ \e^\phi &= \frac{1}{12 \sqrt{4\beta-y\beta^\prime}}\left(\frac{-\beta^\prime}{y}\right)^{5/4}, \label{susysol4} \\ \lambda &= -\frac{216 \pi}{f_0}\left(\frac{-\beta^\prime}{y}\right)^{-3/2}\sqrt{4\beta-y\beta^\prime} \mp \frac{\sqrt{-y\beta^\prime}}{\sqrt{4\beta-y\beta^\prime}}, \label{susysol5}
\end{align}
where the minus (plus) in the last line applies to negative (positive) $y$.
The function $\beta$ satisfies the simple ODE \cite{Apruzzi:2015zna}
\begin{equation}
\beta^{\prime\prime} = \frac{\beta^{\prime 2}}{2\beta} + \frac{\beta^{\prime}}{y} -\frac{f_0\beta^{\prime 3}}{288 \pi y^2 \beta}.
\end{equation}
The general solution is\footnote{Our parameter $c$ is related to $b_2$ in \cite{Apruzzi:2015zna} by $c = 36\sqrt{2}\sqrt{\pi y_0^3/((b_2-18)^3 f_0)}(b_2-12)$.}
\begin{equation}
\sqrt{\beta}= -\frac{4\sqrt{\pi}}{\sqrt{f_0}}\sqrt{y-y_0}(y+2y_0)+c, \label{beta0}
\end{equation}
where $f_0$, $y_0$ and $c$ are free parameters. One can show that $f_0$ and $y_0$ are related to different flux choices \eqref{flux}, while different choices for $c$ correspond to different source distributions (D6-branes, anti-D6-branes or O6-planes) on the two poles of the 3-sphere \cite{Apruzzi:2015zna}.
Furthermore, it is possible to construct even more general solutions with D8-brane stacks by gluing together different solutions \eqref{beta0} at the corresponding brane positions \cite{Apruzzi:2015zna}. In the following, we will focus on solutions without D8-branes.

As a simple example, consider the class of solutions with $c=0$. This corresponds to a configuration where a stack of $N$ D6-branes sits on one pole and the other pole is regular with no sources. We then have
\begin{equation}
\beta = \frac{16\pi}{f_0} (y-y_0)(y+2y_0)^2, \quad y_0 = -\frac{6\pi f_0h^2}{8}, \quad y \in [y_0, -2y_0], \label{beta}
\end{equation}
where the D6-branes are located at $y=-2y_0$. The brane number $N$ is fixed in terms of the flux parameters by the tadpole condition $\frac{1}{2\kappa^2} F_0 \int_{S^3} H_3 = -\mu_6 N$. Using $2\kappa^2=(2\pi)^7\ell_\text{s}^8$, $\mu_6 = (2\pi)^{-6}\ell_\text{s}^{-7}$, one finds
\begin{equation}
f_0 h = -N.
\end{equation}

The equations of motion in the above conventions are \cite{Blaback:2011nz, Blaback:2011pn}
\begin{align}
0 = E_1 &= - \frac{\left[{\e^{-3/2\phi-7A+B+2C} \left({\lambda \e^{7A+3/4\phi}}\right)^\prime}\right]^\prime}{\e^{3B+2C}} + \e^{7/4\phi} \lambda  \frac{f_0^2}{4\pi^2} + \frac{N}{2}\frac{\delta(y+2y_0)}{\e^{3B+2C}}
, \label{vaceom1} \\
0 = E_2 &= -\frac{\left({\e^{7A+B+2C} \, \phi^\prime}\right)^\prime}{\e^{7A+3B+2C}} + \e^{5/2\phi} \frac{f_0^2}{4\pi^2} \left({\frac{5}{4} - \frac{\lambda^2}{2}}\right) + \frac{3}{4} \e^{-14A-2B-3/2\phi} \left({\lambda \e^{7A+3/4\phi}}\right)^{\prime\, 2} \nl + \frac{3N}{8} \frac{\delta(y+2y_0)}{\e^{-3/4\phi+3B+2C}}, \label{vaceom2} \\ 
0 = E_3 &= 96 \e^{-2A} + 16\frac{\left({\e^{7A+B+2C} \, A^\prime}\right)^\prime}{\e^{7A+3B+2C}} + \e^{5/2\phi} \frac{f_0^2}{4\pi^2} \left({1 - 2\lambda^2}\right) - \e^{-14A-2B-3/2\phi} \left({\lambda \e^{7A+3/4\phi}}\right)^{\prime\, 2} \nl - \frac{N}{2} \frac{\delta(y+2y_0)}{\e^{-3/4\phi+3B+2C}}, \\ 
0 = E_4 &= 2 C^{\prime\prime} + 2B^\prime C^\prime + 2 C^{\prime 2} + 7 A^{\prime 2} + 2 B^{\prime\prime} + 7 A^{\prime\prime} - 7 A^\prime B^\prime +\frac{1}{2} \left({\phi^\prime}\right)^2 \nl + \frac{1}{16} \e^{5/2\phi+2B} \frac{f_0^2}{4\pi^2} \left({1+6\lambda^2}\right) - \frac{1}{16} \e^{-14A-3/2\phi} \left({\lambda \e^{7A+3/4\phi}}\right)^{\prime\, 2} + \frac{7N}{32} \frac{\delta(y+2y_0)}{\e^{-3/4\phi+B+2C}}, \\
0 = E_5 &= - \e^{-2C} + C^{\prime\prime} + B^{\prime\prime} + 7A^\prime C^\prime + 7A^\prime B^\prime + 2 C^{\prime 2} + B^{\prime 2} + 3B^\prime C^\prime \nl + \frac{1}{16} \e^{5/2\phi+2B} \frac{f_0^2}{4\pi^2} \left({1+6\lambda^2}\right) + \frac{7}{16} \e^{-14A-3/2\phi} \left({\lambda \e^{7A+3/4\phi}}\right)^{\prime\, 2}  + \frac{7N}{32} \frac{\delta(y+2y_0)}{\e^{-3/4\phi+B+2C}}. \label{vaceom5}
\end{align}
Here, we have only written down the source terms at $y=-2y_0$ that appear in the solution with $c=0$. Also note that the $F_2$ Bianchi identity, which we denoted by $E_1$, is implied by the other equations. We therefore have 4 independent equations for the 4 fields $A$, $B$, $\phi$, $\lambda$ as expected.

The SUSY equations are \cite{Apruzzi:2013yva, Junghans:2014wda}
\begin{align}
0&= S_1= \left(A +\frac{\phi}{4} \right)^\prime - \frac{\e^{-C}}{16} \left[ 4x-\frac{f_0}{2\pi}\e^{A+5/4\phi} \right], \label{susy1} \\
0&= S_2= \phi^\prime - \frac{\e^{-C}}{16} \left[ 12x+ (2x^2-5)\frac{f_0}{2\pi}\e^{A+5/4\phi} \right], \label{susy2} \\
0&= S_3= \left(B+C-A\right)^\prime - \frac{\e^{-C}}{8} \left[ 4x+x^2\frac{f_0}{2\pi}\e^{A+5/4\phi} \right], \\
0&= S_4= \e^{-C}\left[ 6 x + (x^2+x\lambda) \frac{f_0}{2\pi} \e^{A+5/4\phi} \right], \label{susy4}
\end{align}
where $x= \pm \sqrt{1-16\e^{2B+2C-2A}}$. The plus (minus) sign here applies to negative (positive) $y$.

\section{Non-supersymmetric Deformations}
\label{sec:def}

Let us consider a small deformation of the supersymmetric solution \eqref{susysol1}--\eqref{susysol5}, which from now on we label with a subscript ``$0$'':
\begin{align}
\e^{2A(y)} &= \e^{2A_0(y)}+\epsilon a(y), \label{def1} \\ \e^{2B(y)} &= \e^{2B_0(y)}+\epsilon b(y),
\\ \e^{\phi(y)} &= \e^{\phi_0(y)}+\epsilon f(y), \\ \lambda(y) &= \lambda_0(y)+\epsilon l(y). \label{def5}
\end{align}
Note that we do not have to introduce a deformation of $C(y)$ since, as stated above, different choices of $C(y)$ just correspond to a coordinate transformation.

Our goal is now to analyze whether, for given flux numbers $f_0,h$ and fixed sources at the poles, the equations of motion \eqref{vaceom1}--\eqref{vaceom5} can be solved with $a,b,f,l \neq 0$ such that the SUSY equations \eqref{susy1}--\eqref{susy4} are violated at linear order,
\begin{equation}
S_i = \mathcal{O}(\epsilon).
\end{equation}
In general, the direction in which SUSY can be broken (i.e., which combinations of the $S_i$ can be non-vanishing) is determined by the equations of motion. At linear order in $\epsilon$, one finds relations of the form $E_i = a_{ij}S_j^\prime + b_{ij}S_j + \mathcal{O}(\epsilon^2)$, with some $y$-dependent coefficients $a_{ij},b_{ij}$. Specifically, we have (away from the delta-function source terms)
\begin{align}
0=\e^{2B_0} E_2 &= - S_2^\prime + \left[\frac{3}{8}-\frac{3\pi}{x_0f_0}\e^{-A_0-5/4\phi_0}\right]S_4^\prime \nl + \e^{-C}\left[-9x_0+\frac{15}{32}\left(3-4x_0^2\right)\frac{f_0}{\pi}\e^{A_0+5/4\phi_0}+108\frac{\pi}{f_0}\e^{-A_0-5/4\phi_0}\right]S_1 \nl +
\e^{-C}\left[\frac{7}{2}x_0+\frac{1}{32}\left(-7+10x_0^2\right)\frac{f_0}{\pi}\e^{A_0+5/4\phi_0}-36\frac{\pi}{f_0}\e^{-A_0-5/4\phi_0}-\e^C C^\prime\right]S_2 \nl + \e^{-C}\left[-\frac{3}{4x_0}+\frac{3}{32}\left(5-4x_0^2\right)\frac{f_0}{\pi}\e^{A_0+5/4\phi_0}\right]S_3 \nl + \e^{-C}\left[\frac{13}{128}\left(1-x_0^2\right)\frac{f_0}{\pi}\e^{A_0+5/4\phi_0}-\frac{3\pi}{2 x_0^2 f_0}\left(1-x_0^2\right)\e^{-A_0-5/4\phi_0} \right. \nl \qquad\quad \left. -\frac{3\pi}{x_0 f_0} \e^C C^\prime \e^{-A_0-5/4\phi_0} + \frac{3}{8}\e^C C^\prime \right]S_4 + \mathcal{O}(\epsilon^2), \label{eom-susy1}\\
0=\e^{2B_0} E_3 &= 16 S_1^\prime - 4 S_2^\prime + \left[-\frac{1}{2}+\frac{4\pi}{x_0f_0}\e^{-A_0-5/4\phi_0}\right]S_4^\prime \nl
+ \e^{-C}\left[28x_0+\frac{1}{8}\left(17-12x_0^2\right)\frac{f_0}{\pi}\e^{A_0+5/4\phi_0}-144\frac{\pi}{f_0}\e^{-A_0-5/4\phi_0}+16\e^C C^\prime\right]S_1 \nl +
\e^{-C}\left[-10x_0+\frac{1}{8}\left(-3+2x_0^2\right)\frac{f_0}{\pi}\e^{A_0+5/4\phi_0}+48\frac{\pi}{f_0}\e^{-A_0-5/4\phi_0}-4\e^C C^\prime\right]S_2
\nl + \e^{-C}\left[\frac{1}{x_0}+\frac{1}{8}\left(3-4x_0^2\right)\frac{f_0}{\pi}\e^{A_0+5/4\phi_0}\right]S_3 
\nl + \e^{-C}\left[\frac{2}{x_0}(1-x_0^2)+\frac{1}{32}\left(1-x_0^2\right)\frac{f_0}{\pi}\e^{A_0+5/4\phi_0}+\frac{2\pi}{x_0^2 f_0}\left(1-x_0^2\right)\e^{-A_0-5/4\phi_0} \right. \nl \qquad\quad \left. +\frac{4\pi}{x_0 f_0} \e^C C^\prime \e^{-A_0-5/4\phi_0} - \frac{1}{2}\e^C C^\prime \right]S_4 + \mathcal{O}(\epsilon^2), \\
0=E_4 &= 9 S_1^\prime- \frac{9}{4} S_2^\prime + 2 S_3^\prime + \left[-\frac{1}{32}+\frac{\pi}{4x_0f_0}\e^{-A_0-5/4\phi_0}\right]S_4^\prime \nl
+ \e^{-C}\left[-\frac{13}{4}x_0+\frac{1}{128}\left(9-44x_0^2\right)\frac{f_0}{\pi}\e^{A_0+5/4\phi_0}-9\frac{\pi}{f_0}\e^{-A_0-5/4\phi_0}+9\e^C C^\prime\right]S_1
\nl + \e^{-C}\left[\frac{11}{8}x_0+\frac{1}{128}\left(-11+18x_0^2\right)\frac{f_0}{\pi}\e^{A_0+5/4\phi_0}+3\frac{\pi}{f_0}\e^{-A_0-5/4\phi_0}-\frac{9}{4}\e^C C^\prime\right]S_2
\nl + \e^{-C}\left[\frac{1}{16x_0}(-23+16x_0^2)+\frac{1}{128}\left(-5+12x_0^2\right)\frac{f_0}{\pi}\e^{A_0+5/4\phi_0} + 2\e^C C^\prime\right]S_3 
\nl + \e^{-C}\left[-\frac{1}{4 x_0}(1-x_0^2)-\frac{15}{512}\left(1-x_0^2\right)\frac{f_0}{\pi}\e^{A_0+5/4\phi_0}+\frac{\pi}{8 x_0^2f_0}\left(1-x_0^2\right)\e^{-A_0-5/4\phi_0} \right. \nl \qquad\quad \left. +\frac{\pi}{4 x_0 f_0} \e^C C^\prime \e^{-A_0-5/4\phi_0} - \frac{1}{32}\e^C C^\prime \right]S_4 + \mathcal{O}(\epsilon^2), \\
0=E_5 &= S_1^\prime- \frac{1}{4} S_2^\prime + S_3^\prime + \left[\frac{7}{32}-\frac{7\pi}{4x_0f_0}\e^{-A_0-5/4\phi_0}\right]S_4^\prime \nl
+ \e^{-C}\left[\frac{15}{4}x_0+\frac{1}{128}\left(17-52x_0^2\right)\frac{f_0}{\pi}\e^{A_0+5/4\phi_0}+63\frac{\pi}{f_0}\e^{-A_0-5/4\phi_0}+\e^C C^\prime\right]S_1
\nl + \e^{-C}\left[\frac{3}{8}x_0+\frac{1}{128}\left(-3+10x_0^2\right)\frac{f_0}{\pi}\e^{A_0+5/4\phi_0}-21\frac{\pi}{f_0}\e^{-A_0-5/4\phi_0}-\frac{1}{4}\e^C C^\prime\right]S_2
\nl + \e^{-C}\left[\frac{1}{16x_0}(-23+48x_0^2)+\frac{1}{128}\left(11-4x_0^2\right)\frac{f_0}{\pi}\e^{A_0+5/4\phi_0}+\e^C C^\prime\right]S_3 
\nl + \e^{-C}\left[-\frac{1}{2x_0}(1-x_0^2)+\frac{9}{512}\left(1-x_0^2\right)\frac{f_0}{\pi}\e^{A_0+5/4\phi_0}-\frac{7\pi}{8 x_0^2 f_0}\left(1-x_0^2\right)\e^{-A_0-5/4\phi_0} \right. \nl \qquad\quad \left. -\frac{7\pi}{4 x_0 f_0} \e^C C^\prime \e^{-A_0-5/4\phi_0} + \frac{7}{32}\e^C C^\prime \right]S_4 + \mathcal{O}(\epsilon^2). \label{eom-susy4}
\end{align}
In order to obtain these relations, we have replaced $A^\prime$, $A^{\prime\prime}$, $B^\prime$, $B^{\prime\prime}$, $\phi^\prime$, $\phi^{\prime\prime}$, $\lambda$, $\lambda^\prime$ in \eqref{vaceom2}--\eqref{vaceom5} by $S_i$, $S_i^\prime$ using \eqref{susy1}--\eqref{susy4}. The result is a system of 4 coupled 1st-order ODEs for the unknown functions $S_i(y)$ with some known $y$-dependent coefficients that are given in terms of the supersymmetric solution.

The ODE system \eqref{eom-susy1}--\eqref{eom-susy4} can be significantly simplified by defining
\begin{equation}
\frac{1}{4\pi}S_5 = 12 S_1 - 4 S_2 - \frac{1}{2} S_4. \label{s5-def}
\end{equation}
A key observation is now that, in terms of $S_5$, $S_1$, $S_2$, $S_3$, the system \eqref{eom-susy1}--\eqref{eom-susy4} decouples. In particular, it is now possible to find linear combinations of \eqref{eom-susy1}--\eqref{eom-susy4} such that the ODE system takes the following form:
\begin{align}
0 &= S_5^{\prime\prime} + c_{51} S_5^\prime + c_{52} S_5, \label{ode-s5-gen} \\
0 &= S_1^{\prime\prime} + c_{11} S_1^\prime + c_{12} S_1 + c_{13} S_5^\prime + c_{14} S_5, \label{ode-s1-gen} \\
0 &= S_2 + c_{21} S_1^\prime + c_{22} S_1 + c_{23} S_5^\prime + c_{24} S_5, \label{ode-s2-gen} \\
0 &= S_3 + c_{31} S_1 + c_{32}S_2 + c_{33}S_5^\prime + c_{34}S_5 \label{ode-s3-gen}
\end{align}
with some $y$-dependent coefficients $c_{ij}$ which again depend on the supersymmetric solution (i.e., on $A_0(y),B_0(y),C(y),\phi_0(y)$).
This ODE system can be derived as follows: First, consider linear combinations of \eqref{eom-susy1}--\eqref{eom-susy4} such that their dependence on $S_2^\prime$ and $S_3^\prime$ cancels out. This yields the two linearly independent equations \eqref{ode-s2-gen} and \eqref{ode-s3-gen}, which can be solved for $S_2$ and $S_3$. Plugging back the solution into \eqref{eom-susy1}--\eqref{eom-susy4} then results in two further linearly independent equations that only depend on $S_1$ and $S_5$ as well as their first and second derivatives. We can now find a linear combination of these two equations such that its dependence on $S_1$, $S_1^\prime$ and $S_1^{\prime\prime}$ vanishes, which yields \eqref{ode-s5-gen}. The remaining equation is then \eqref{ode-s1-gen}.

The equations \eqref{ode-s5-gen}--\eqref{ode-s3-gen} can now be solved one after another. We will argue in the following sections that the only consistent solution to the first ODE is to choose $S_5=0$. For this choice, the second ODE only depends on $S_1$, and we will again argue that the only consistent solution is $S_1=0$. It then follows from the last two equations that also $S_2=S_3=0$, showing that there are no deformations of the supersymmetric solution with $S_i \neq 0$ as claimed.\footnote{We stress that it is not a limitation of our result that we performed our analysis at linear order in $\epsilon$. Indeed, showing an obstruction to a flat potential/a modulus already at linear order is sufficient to rule out its existence, and no further analysis at higher orders is required.}

\section{ODE for $S_5(y)$}
\label{sec:ode}

Let us now consider the linear 2nd-order ODE for $S_5$ \eqref{ode-s5-gen}. We will assume that $S_5$ is a smooth function away from the poles where the branes sit.\footnote{This is perhaps not obvious given that the definition of the $S_i$ in \eqref{susy1}--\eqref{susy4} involves the piecewise-defined function $x$. One may therefore wonder whether the $S_i$ or their derivatives may have discontinuities even though $A$, $B$, $C$, $\phi$, $\lambda$ have to be smooth away from the branes. To see that this is not the case, consider the linear combinations $T_1 =S_1-\frac{1}{3}S_2$, $T_2 =S_1-\frac{1}{2}S_3$ and $T_3 =S_1+\frac{1}{6}\lambda\frac{f_0}{2\pi}\e^{A+5/4\phi}S_1+\frac{1}{24}S_4$, which only depend on $x^2$ but not on $x$ itself and must therefore be smooth. Rewriting \eqref{eom-susy1}--\eqref{eom-susy4} in terms of $S_1$, $T_1$, $T_2$ and $T_3$, one can derive an expression for $S_1$ that only involves the smooth functions $T_1$, $T_2$, $T_3$ and their derivatives at linear order in $\epsilon$. The coefficients in front of $S_1$ and the $T_i$ are also smooth since $x_0$ can be expressed in terms of smooth functions using, e.g., \eqref{susy1}.
Since $S_1$ is thus given by a sum of smooth functions, it must itself be smooth. Using the above definitions of $T_1$, $T_2$ and $T_3$, it then follows that also $S_2$, $S_3$ and $S_4$ must be smooth.} For convenience, we furthermore define $s_5 = \beta S_5$ and use \eqref{susysol1}--\eqref{susysol4}, \eqref{beta0} to simplify the coefficients of the ODE. This yields the equation
\begin{equation}
s_5(y)^{\prime\prime}+p(y)s_5(y)^{\prime}+q(y)s_5(y)=0 \label{ode-s5}
\end{equation}
with coefficients
\begin{align}
p(y) &=\frac{1}{2}\frac{9\sqrt{\pi f_0}y^2+2f_0\sqrt{\beta(y)}\sqrt{y-y_0}}{3\sqrt{\pi f_0}y^2(y-y_0)+f_0\sqrt{\beta(y)}(y-y_0)^{3/2}}, \\ q(y) &= \frac{432\pi y^2(y-y_0)+24\sqrt{\pi f_0}\sqrt{\beta(y)}\sqrt{y-y_0}(5y-6y_0)-f_0\beta(y)}{4\sqrt{\beta(y)}\left[3\sqrt{\pi f_0}y^2(y-y_0)^{3/2}+f_0\sqrt{\beta(y)}(y-y_0)^2\right]}.
\end{align}

To be explicit, let us now consider the simple solution with D6-branes on one pole for which $\beta$ is given by \eqref{beta}. For simplicity, we furthermore set $f_0=1$, $y_0=-1$. We can do this without loss of generality since the solution with this flux choice can be mapped to a solution with any other flux choice by a simple rescaling of $A,B,\phi,\lambda$ (see, e.g., \cite{Junghans:2014wda}). The above expressions thus simplify to
\begin{align}
p(y) =-\frac{1}{2} \frac{y^2+8y+16}{y^3-3y^2-12y-8}, \quad q(y) = \frac{-4y^2+28y+68}{y^4-5y^3-6y^2+16y+16}. \label{ode-s5-coeffs}
\end{align}
We observe that the coefficients have simple poles at $y=\hat y_i$ with
\begin{equation}
\hat y_1 = -1, \quad \hat y_2=2
\end{equation}
(as well two other poles at $y\approx 5.46$, $y\approx -1.46$ which lie outside of the domain of definition of $y\in [-1,2]$).

Closed-form solutions to linear 2nd-order ODEs are only known if the coefficients are constant or their $y$-dependence takes a special form such as, e.g., in the hypergeometric DE. Consequently, an algorithm that solves a general linear 2nd-order ODE with arbitrary $y$-dependent coefficients is not available. Indeed, one can check using algorithms provided in computer algebra systems that there is no Liouvillian or hypergeometric function satisfying the above ODE.
However, a quite general way to construct a solution to an ODE of the above type is in terms of an infinite series around some point $y=\hat y$.
This is called the Frobenius method, and there are some useful math results regarding the existence and properties of such solutions.

Fuchs's theorem states that a solution to the ODE always exists around $y=\hat y$ provided that $y=\hat y$ is either a non-singular point or a regular singular point, i.e., the coefficients $p(y),q(y)$ only have poles of at most the orders $p(y)\sim (y-\hat y)^{-1}$, $q(y)\sim (y-\hat y)^{-2}$. Since the coefficients \eqref{ode-s5-coeffs} only have simple poles, this is indeed the case for every point on the internal space. Furthermore, the convergence radius of the Frobenius series is at least as large as the minimum of the convergence radii of the corresponding expansions of $p(y)(y-\hat y)$ and $q(y)(y-\hat y)^2$. We can easily test the convergence radii $r$ of $p(y)(y-\hat y),q(y)(y-\hat y)^2$ at any point by substituting their expansions into $r=\lim_{n\to\infty}|\frac{c_n}{c_{n+1}}|$, where $c_n$ are the corresponding expansion coefficients.

Hence, in order to construct the full solution for $s_5(y)$, we can pick a few points $\hat y$ on the interval $y\in[-1,2]$ where we construct Frobenius solutions $s_5(y;\hat y)$ such that their domains of validity overlap. These solutions can then be glued together at some intermediate points such that we arrive at a solution for $s_5(y)$ that is well-defined over the whole interval. At each point, there are two linearly independent solutions for $s_5(y)$, where the general solution is an arbitrary linear combination of them. The overall scale of $s_5(y)$ drops out of the ODE. The general solution therefore has one free parameter.

Let us now state the solution. Around $y=2$, the general solution is
\begin{equation}
s_5(y;2) = \alpha_2 s_5^{(1)}(y;2) + \beta_2 s_5^{(2)}(y;2)
\end{equation}
with arbitrary coefficients $\alpha_2,\beta_2$ and
\begin{align}
s_5^{(1)}(y;2) &= 1 - \frac{161}{24} (2-y)^{2} + \frac{1405}{432} (2-y)^{3} - \frac{433}{5184} (2-y)^{4} \nl
+\ln(2-y)\left[-3(2-y)+\frac{15}{4}(2-y)^2-\frac{11}{12}(2-y)^3-\frac{5}{288}(2-y)^4\right] \nl
+ \mathcal{O}((2-y)^5), \label{s5-frob1} \\
s_5^{(2)}(y;2) &= (2-y) - \frac{5}{4} (2-y)^{2} + \frac{11}{36}(2-y)^{3} + \frac{5}{864}(2-y)^{4} + \mathcal{O}((2-y)^5). \label{s5-frob2}
\end{align}

Also note that, at zeroth order in the $\epsilon$-expansion, the terms in \eqref{susy1}--\eqref{susy4} diverge at most like $(2-y)^{-1}$. This can easily be verified by substituting the SUSY solution and expanding in $2-y$. Since, by assumption, our $\mathcal{O}(\epsilon)$ deformation is a small correction to the leading-order terms everywhere on the internal space, it must not produce terms $\sim (2-y)^{-2}$ in $S_1$, $S_2$, $S_3$, $S_4$. We should therefore discard the too singular solution $s_5^{(1)} \sim (2-y)^{0}$ (which would yield $S_5 \sim (2-y)^{-2}$) and only consider the solution $s_5^{(2)} \sim (2-y)^{1}$. Another argument leading to the same conclusion is that the most general behavior of the fields at the poles that is compatible with the equations of motion was already computed in \cite{Blaback:2011pn}. Using this result in \eqref{susy1}--\eqref{susy4} again shows that the terms in these equations cannot diverge faster than $(2-y)^{-1}$. Hence, a self-consistent deformation requires the boundary condition $\alpha_2=0$ at $y=2$.

At $y=-1$, we find
\begin{equation}
s_5(y;-1) = \alpha_{-1} s_5^{(1)}(y;-1) + \beta_{-1} s_5^{(2)}(y;-1)
\end{equation}
with arbitrary coefficients $\alpha_{-1},\beta_{-1}$ and
\begin{align}
s_5^{(1)}(y;-1) &= (y+1)^{-1/2} - 10(y+1)^{1/2} + \frac{35}{3} (y+1)^{3/2} - \frac{56}{15}(y+1)^{5/2} + \frac{22}{105}(y+1)^{7/2} \nl + \mathcal{O}((y+1)^{9/2}),  \label{s5-frob3} \\
s_5^{(2)}(y;-1) &= 1-\frac{8}{3}(y+1) + \frac{8}{5}(y+1)^{2} - \frac{16}{63}(y+1)^{3} -\frac{2}{567}(y+1)^4 + \mathcal{O}((y+1)^5).  \label{s5-frob4}
\end{align}
At zeroth order in the $\epsilon$-expansion, the terms in \eqref{susy1} diverge at most like $(y+1)^{-1}$. The solution $s_5^{(1)}\sim (y+1)^{-1/2}$ would yield $S_5\sim (y+1)^{-3/2}$ and is therefore not consistent with our $\epsilon$-expansion. Hence, analogously as for the other pole, only $s_5^{(2)}(y)$ is a valid solution. We therefore require $\alpha_{-1}=0$ as a boundary condition at $y=-1$.

\begin{figure}[t]
\centering
\includegraphics[trim = 0mm 0mm 0mm 0mm, width=0.4\textwidth]{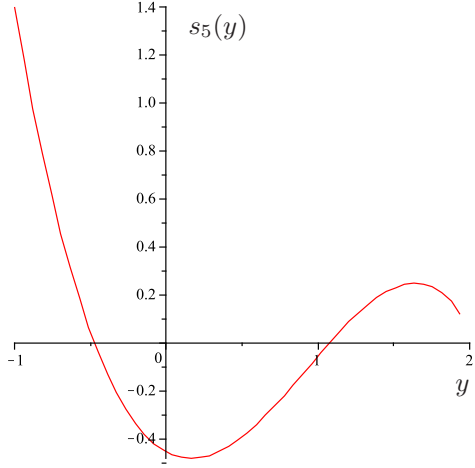}
\put(-110,165){\footnotesize{$s_5(y)$}}
\put(-10,30){\footnotesize{$y$}}
\caption{Plot of $s_5(y)$ between the poles $y=-1$ and $y=2$.} \label{fig1}
\end{figure}

Computing the convergence radii of the expansions of $p(y)(y-\hat y_i)$, $q(y)(y-\hat y_i)^2$ at $y=2$ and $y=-1$, we find that the convergence radius of $s_5(y;2)$ is at least $3$ and the convergence radius of $s_5(y;-1)$ is at least $\approx 0.46$. Hence, the convergence radii of the two solutions overlap and we may match the Frobenius series at, e.g., $y=-0.7$. Doing this, we find
\begin{equation}
\alpha_{-1} \approx 0, \qquad \beta_{-1} \approx 1.4,\qquad \alpha_2 \approx 0.047, \qquad \beta_2=1
\end{equation}
up to an overall rescaling.\footnote{Note that the small non-vanishing value for $\alpha_2$ is not a numerical artifact. If one goes to higher orders in the Frobenius expansion, the values for $\alpha_2$ remain stable and non-vanishing. We have tested this including terms up to 10th order.} Hence, only one globally consistent solution with the allowed boundary conditions at $y=-1$ ($\alpha_{-1}=0$) exists (see Fig.\ \ref{fig1}). It leads to a too singular field behavior at $y=2$ ($\alpha_{2}\neq 0$). Similarly, one can show that a solution with $\alpha_2 =0$ is matched to a solution with a too singular behavior at $y=-1$ ($\alpha_{-1}\neq 0$).
Since we excluded the boundary conditions with $\alpha_{-1},\alpha_{2}\neq 0$ above, it follows that we have to set
\begin{equation}
S_5(y)=0. \label{no-s5}
\end{equation}

As stated above, we restricted our discussion here to deformations of the solution with one stack of D6-branes (corresponding to the choice $c=0$ in \eqref{beta0}). We also checked several examples of solutions with sources at both poles (i.e., $c \neq 0$) and found that analogous arguments again imply \eqref{no-s5}.\footnote{For cases involving an O6-plane (i.e., $b_2> 12$ in the conventions of \cite{Apruzzi:2015zna}), the argument is more subtle than for cases with only D-branes ($b_2 \le 12$) because the solutions of \cite{Apruzzi:2015zna} are only defined up to the boundary of the hole created by the O6-plane at $y=-1$. Our previous argument involving the local behavior at the source position does therefore not directly apply for such solutions. However, one can check that a non-zero $s_5$ deformation would either have a too singular boundary condition at the pole opposite to the O6 or introduce new singular terms in the fields at the hole boundary. If the latter is not admissible, such deformations are then ruled out as well.} However, although the examples we checked do not indicate it, we cannot exclude that there are special values of $c\neq 0$ for which the matching does work.

More generally, it is possible to construct solutions with one or several D8-brane stacks inbetween the poles \cite{Apruzzi:2013yva, Junghans:2014wda, Apruzzi:2015zna}. We have not studied such solutions in detail. The presence of the D8-branes would lead to modified matching conditions at the positions of the stacks and thus significantly complicate the analysis (in particular for solutions with many stacks). Nevertheless, it is possible to study such solutions using the same techniques that we discussed above. We leave this task for future work.

\section{ODE for $S_1(y)$}
\label{sec:ode2}

We now proceed with solving the ODE system \eqref{ode-s5-gen}--\eqref{ode-s3-gen}. As we have just shown, the only solution to the first of these equations is $S_5=0$. Using this, \eqref{ode-s1-gen} becomes a linear 2nd-order ODE that only depends on $S_1$. It is convenient to define $s_1 = y(y-y_0)S_1$ to further simplify the equation. We thus find
\begin{equation}
s_1(y)^{\prime\prime}+p(y)s_1(y)^{\prime}+q(y)s_1(y)=0 \label{ode}
\end{equation}
with coefficients
\begin{align}
p(y) &= \frac{222y^5-440y^4-2755y^3-3798y^2-1112y+512}{148y^6-620y^5-410y^4+1706y^3+1156y^2-512y-320}, \label{ode-coeffs1} \\ q(y) &= \frac{222y^4-1664y^3-480y^2+3525y+2200}{74y^6-310y^5-205y^4+853y^3+578y^2-256y-160}, \label{ode-coeffs2}
\end{align}
where we again specialize to the solution with $c=0$, $f_0=1$, $y_0=-1$.
The coefficients have simple poles at $y=\hat y_i$ with
\begin{equation}
\hat y_1 = -1, \quad  \hat y_2\approx -0.971, \quad \hat y_3\approx -0.482, \quad \hat y_4\approx 0.568, \quad \hat y_5=2
\end{equation}
(as well as another one at $y\approx 4.07$ which lies outside of the domain of definition of $y\in [-1,2]$). The poles are therefore regular singular points.

As for the ODE for $s_5$ we had above, we can again check here that there is no Liouvillian or hypergeometric function that solves the ODE. We therefore choose again to construct the solution in terms of a Frobenius series. At $y=2$, we have the two solutions
\begin{align}
s_1^{(1)}(y;2) &= -\frac{1080}{7} (2-y)^{-3} + 144 (2-y)^{-2} - \frac{271}{7} (2-y)^{-1} + 8\ln(2-y) + \frac{121}{2} \nl + \mathcal{O}(2-y), \label{frob1} \\
s_1^{(2)}(y;2) &= 8 + (2-y) - \frac{4}{3} (2-y)^{2} - \frac{79}{120}(2-y)^{3} - \frac{9}{40}(2-y)^{4} + \mathcal{O}((2-y)^5) \label{frob2}
\end{align}
up to overall rescalings of the functions. At zeroth order in the $\epsilon$-expansion, the terms in \eqref{susy1} diverge at most like $(2-y)^{-1}$. Since, by assumption, our $\mathcal{O}(\epsilon)$ deformation is a small correction to the LO terms, it can then not produce terms $\sim (2-y)^{-3}$ in $S_1$. We should therefore discard the singular solution $s_1^{(1)} \sim (2-y)^{-3}$, and only consider the regular solution $s_1^{(2)} \sim (2-y)^{0}$.

At $y=-1$, we find
\begin{align}
s_1^{(1)}(y;-1) &= (y+1)^{-1/2} + \frac{124}{3}(y+1)^{1/2} + \frac{89}{9} (y+1)^{3/2} - \frac{5344}{135}(y+1)^{5/2} \nl - \frac{18799}{405}(y+1)^{7/2} + \mathcal{O}((y+1)^{9/2}),  \label{frob3} \\
s_1^{(2)}(y;-1) &= 1+2(y+1) - \frac{16}{15} (y+1)^{2} - \frac{38}{15}(y+1)^{3} - \frac{98}{45}(y+1)^{4} + \mathcal{O}((y+1)^5).  \label{frob4}
\end{align}
At zeroth order in the $\epsilon$-expansion, the terms in \eqref{susy1} diverge at most like $(y+1)^{-1}$. The solution $s_1^{(1)}\sim (y+1)^{-1/2}$ would yield $S_1\sim (y+1)^{-3/2}$ and is therefore not consistent with our $\epsilon$-expansion. Hence, again only $s_1^{(2)}(y)$ is a valid solution.

Similarly, we can compute solutions in terms of Frobenius series at the other singular points $y=\hat y_2,\hat y_3,\hat y_4$ (or at any other point on the sphere). Gluing together these solutions at some intermediate points (as explained in section \ref{sec:ode}), we find that it is not possible to choose linear combinations of them such that a globally consistent solution is obtained. We thus conclude that we have to set
\begin{equation}
S_1(y)= 0.
\end{equation}
The remaining functions are $S_2$ and $S_3$. According to \eqref{ode-s2-gen}, \eqref{ode-s3-gen}, they are linear combinations of $S_5$, $S_5^\prime$, $S_1$ and $S_1^\prime$. It follows that
\begin{equation}
S_2(y)=S_3(y)=0.
\end{equation}
Hence, we have shown that there are no supersymmetry-breaking deformations of the supersymmetric solution.

As in section \ref{sec:ode}, we restricted our discussion here to the simple case $c=0$. However, we also studied more complicated solutions with branes/O-planes on both poles (i.e., with $c\neq 0$), again with the same result.\footnote{The subtlety discussed in footnote 9 for the O6-plane case does not apply here because the matching of the Frobenius series for $s_1$ leads to an inconsistency regardless of the local behavior near the O6-plane hole boundary. In particular, the matching reveals that $S_1$ either diverges at $y=0$ or has a too singular boundary condition at the pole opposite to the O6.}

\section{Conclusions}
\label{sec:concl}

In this paper, we have shown that a simple class of supersymmetric AdS$_7$ vacua of massive type IIA string theory does not admit supersymmetry-breaking deformations. Our result complements previous analyses in \cite{Louis:2015mka, Cordova:2016xhm} which showed that no supersymmetric deformations are possible. It follows that the supersymmetric AdS$_7$ vacua we studied have no moduli space. Via holography, this suggests that the dual 6D $(1,0)$ theories do not have a conformal manifold.

In our analysis, we focussed on AdS solutions with (anti-)D6-branes and/or O6-planes. More generally, one can consider solutions with one or several D8-brane stacks \cite{Apruzzi:2013yva, Junghans:2014wda, Apruzzi:2015zna}, and it would be interesting to extend our analysis to such setups. It would also be interesting to further study non-supersymmetric AdS solutions in this setup that are not continuously connected to a supersymmetric one, in particular in view of the conjecture in \cite{Ooguri:2016pdq}.
We hope to come back to some of these open problems in future work.

\section*{Acknowledgements}

We would like to thank Fabio Apruzzi, Severin L\"{u}st and Stefano Massai for useful discussions.

\bibliographystyle{utphys}
\bibliography{groups}

\providecommand{\href}[2]{#2}\begingroup\raggedright\begin{thebibliography}{10}

\bibitem{Kachru:2003aw}
S.~Kachru, R.~Kallosh, A.~D. Linde and S.~P. Trivedi,  {\em {De Sitter vacua in
  string theory}}, Phys. Rev. {\bf D68} (2003) 046005
[\href{http://www.arXiv.org/abs/hep-th/0301240}{{\tt hep-th/0301240}}].

\bibitem{Balasubramanian:2005zx}
V.~Balasubramanian, P.~Berglund, J.~P. Conlon and F.~Quevedo,  {\em
  {Systematics of moduli stabilisation in Calabi-Yau flux compactifications}},
  JHEP {\bf 03} (2005) 007
[\href{http://www.arXiv.org/abs/hep-th/0502058}{{\tt hep-th/0502058}}].

\bibitem{Maldacena:1997re}
J.~M. Maldacena,  {\em {The Large N limit of superconformal field theories and
  supergravity}}, Int. J. Theor. Phys. {\bf 38} (1999) 1113--1133
  [\href{http://www.arXiv.org/abs/hep-th/9711200}{{\tt hep-th/9711200}}],
[Adv. Theor. Math. Phys.2,231(1998)].

\bibitem{Gubser:1998bc}
S.~S. Gubser, I.~R. Klebanov and A.~M. Polyakov,  {\em {Gauge theory
  correlators from noncritical string theory}}, Phys. Lett. {\bf B428} (1998)
  105--114
[\href{http://www.arXiv.org/abs/hep-th/9802109}{{\tt hep-th/9802109}}].

\bibitem{Witten:1998qj}
E.~Witten,  {\em {Anti-de Sitter space and holography}}, Adv. Theor. Math.
  Phys. {\bf 2} (1998) 253--291
[\href{http://www.arXiv.org/abs/hep-th/9802150}{{\tt hep-th/9802150}}].

\bibitem{Blaback:2011nz}
J.~Bl{\r{a}}b{\"{a}}ck, U.~H. Danielsson, D.~Junghans, T.~Van~Riet, T.~Wrase
  and M.~Zagermann,  {\em {The problematic backreaction of SUSY-breaking
  branes}}, JHEP {\bf 08} (2011) 105
[\href{http://www.arXiv.org/abs/1105.4879}{{\tt 1105.4879}}].

\bibitem{Blaback:2011pn}
J.~Bl{\r{a}}b{\"{a}}ck, U.~H. Danielsson, D.~Junghans, T.~Van~Riet, T.~Wrase
  and M.~Zagermann,  {\em {(Anti-)Brane backreaction beyond perturbation
  theory}}, JHEP {\bf 1202} (2012) 025
[\href{http://www.arXiv.org/abs/1111.2605}{{\tt 1111.2605}}].

\bibitem{Blaback:2010sj}
J.~Bl{\r{a}}b{\"{a}}ck, U.~H. Danielsson, D.~Junghans, T.~Van~Riet, T.~Wrase
  and M.~Zagermann,  {\em {Smeared versus localised sources in flux
  compactifications}}, JHEP {\bf 12} (2010) 043
[\href{http://www.arXiv.org/abs/1009.1877}{{\tt 1009.1877}}].

\bibitem{Danielsson:2013qfa}
U.~H. Danielsson, G.~Dibitetto, M.~Fazzi and T.~Van~Riet,  {\em {A note on
  smeared branes in flux vacua and gauged supergravity}}, JHEP {\bf 04} (2014)
  025
[\href{http://www.arXiv.org/abs/1311.6470}{{\tt 1311.6470}}].

\bibitem{Apruzzi:2013yva}
F.~Apruzzi, M.~Fazzi, D.~Rosa and A.~Tomasiello,  {\em {All AdS\_7 solutions of
  type II supergravity}}, JHEP {\bf 04} (2014) 064
[\href{http://www.arXiv.org/abs/1309.2949}{{\tt 1309.2949}}].

\bibitem{Junghans:2014wda}
D.~Junghans, D.~Schmidt and M.~Zagermann,  {\em {Curvature-induced Resolution
  of Anti-brane Singularities}}, JHEP {\bf 10} (2014) 34
[\href{http://www.arXiv.org/abs/1402.6040}{{\tt 1402.6040}}].

\bibitem{Apruzzi:2015zna}
F.~Apruzzi, M.~Fazzi, A.~Passias and A.~Tomasiello,  {\em {Supersymmetric
  AdS$\_{5}$ solutions of massive IIA supergravity}}, JHEP {\bf 06} (2015) 195
[\href{http://www.arXiv.org/abs/1502.06620}{{\tt 1502.06620}}].

\bibitem{Gaiotto:2014lca}
D.~Gaiotto and A.~Tomasiello,  {\em {Holography for (1,0) theories in six
  dimensions}}, JHEP {\bf 12} (2014) 003
[\href{http://www.arXiv.org/abs/1404.0711}{{\tt 1404.0711}}].

\bibitem{Cremonesi:2015bld}
S.~Cremonesi and A.~Tomasiello,  {\em {6d holographic anomaly match as a
  continuum limit}}, JHEP {\bf 05} (2016) 031
[\href{http://www.arXiv.org/abs/1512.02225}{{\tt 1512.02225}}].

\bibitem{Apruzzi:2017nck}
F.~Apruzzi and M.~Fazzi,  {\em {AdS$_{7}$/CFT$_{6}$ with orientifolds}}, JHEP
  {\bf 01} (2018) 124
[\href{http://www.arXiv.org/abs/1712.03235}{{\tt 1712.03235}}].

\bibitem{DeLuca:2018zbi}
G.~B. De~Luca, A.~Gnecchi, G.~Lo~Monaco and A.~Tomasiello,  {\em {Holographic
  duals of 6d RG flows}},
\href{http://www.arXiv.org/abs/1810.10013}{{\tt 1810.10013}}.

\bibitem{Intriligator:1997kq}
K.~A. Intriligator,  {\em {RG fixed points in six-dimensions via branes at
  orbifold singularities}}, Nucl. Phys. {\bf B496} (1997) 177--190
[\href{http://www.arXiv.org/abs/hep-th/9702038}{{\tt hep-th/9702038}}].

\bibitem{Intriligator:1997dh}
K.~A. Intriligator,  {\em {New string theories in six-dimensions via branes at
  orbifold singularities}}, Adv. Theor. Math. Phys. {\bf 1} (1998) 271--282
[\href{http://www.arXiv.org/abs/hep-th/9708117}{{\tt hep-th/9708117}}].

\bibitem{Brunner:1997gf}
I.~Brunner and A.~Karch,  {\em {Branes at orbifolds versus Hanany Witten in
  six-dimensions}}, JHEP {\bf 03} (1998) 003
[\href{http://www.arXiv.org/abs/hep-th/9712143}{{\tt hep-th/9712143}}].

\bibitem{Hanany:1997gh}
A.~Hanany and A.~Zaffaroni,  {\em {Branes and six-dimensional supersymmetric
  theories}}, Nucl. Phys. {\bf B529} (1998) 180--206
[\href{http://www.arXiv.org/abs/hep-th/9712145}{{\tt hep-th/9712145}}].

\bibitem{Janssen:1999sa}
B.~Janssen, P.~Meessen and T.~Ortin,  {\em {The D8-brane tied up: String and
  brane solutions in massive type IIA supergravity}}, Phys. Lett. {\bf B453}
  (1999) 229--236
[\href{http://www.arXiv.org/abs/hep-th/9901078}{{\tt hep-th/9901078}}].

\bibitem{Imamura:2001cr}
Y.~Imamura,  {\em {1/4 BPS solutions in massive IIA supergravity}}, Prog.
  Theor. Phys. {\bf 106} (2001) 653--670
[\href{http://www.arXiv.org/abs/hep-th/0105263}{{\tt hep-th/0105263}}].

\bibitem{Bobev:2016phc}
N.~Bobev, G.~Dibitetto, F.~F. Gautason and B.~Truijen,  {\em {Holography, Brane
  Intersections and Six-dimensional SCFTs}}, JHEP {\bf 02} (2017) 116
[\href{http://www.arXiv.org/abs/1612.06324}{{\tt 1612.06324}}].

\bibitem{Macpherson:2016xwk}
N.~T. Macpherson and A.~Tomasiello,  {\em {Minimal flux Minkowski
  classification}}, JHEP {\bf 09} (2017) 126
[\href{http://www.arXiv.org/abs/1612.06885}{{\tt 1612.06885}}].

\bibitem{Passias:2015gya}
A.~Passias, A.~Rota and A.~Tomasiello,  {\em {Universal consistent truncation
  for 6d/7d gauge/gravity duals}},
\href{http://www.arXiv.org/abs/1506.05462}{{\tt 1506.05462}}.

\bibitem{Malek:2018zcz}
E.~Malek, H.~Samtleben and V.~Vall~Camell,  {\em {Supersymmetric AdS$_{7}$ and
  AdS$_6$ vacua and their minimal consistent truncations from exceptional field
  theory}}, Phys. Lett. {\bf B786} (2018) 171--179
[\href{http://www.arXiv.org/abs/1808.05597}{{\tt 1808.05597}}].

\bibitem{Malek:2019ucd}
E.~Malek, H.~Samtleben and V.~Vall~Camell,  {\em {Supersymmetric AdS$_7$ and
  AdS$_6$ vacua and their consistent truncations with vector multiplets}},
\href{http://www.arXiv.org/abs/1901.11039}{{\tt 1901.11039}}.

\bibitem{Dibitetto:2015bia}
G.~Dibitetto, J.~J. Fernandez-Melgarejo and D.~Marques,  {\em {All gaugings and
  stable de Sitter in D = 7 half-maximal supergravity}}, JHEP {\bf 11} (2015)
  037
[\href{http://www.arXiv.org/abs/1506.01294}{{\tt 1506.01294}}].

\bibitem{Apruzzi:2016rny}
F.~Apruzzi, G.~Dibitetto and L.~Tizzano,  {\em {A new 6d fixed point from
  holography}},
\href{http://www.arXiv.org/abs/1603.06576}{{\tt 1603.06576}}.

\bibitem{Dine:1987vf}
M.~Dine and N.~Seiberg,  {\em {Microscopic Knowledge From Macroscopic Physics
  in String Theory}}, Nucl. Phys. {\bf B301} (1988)
357--380.

\bibitem{Banks:1988yz}
T.~Banks and L.~J. Dixon,  {\em {Constraints on String Vacua with Space-Time
  Supersymmetry}}, Nucl. Phys. {\bf B307} (1988)
93--108.

\bibitem{Bashmakov:2017rko}
V.~Bashmakov, M.~Bertolini and H.~Raj,  {\em {On non-supersymmetric conformal
  manifolds: field theory and holography}}, JHEP {\bf 11} (2017) 167
[\href{http://www.arXiv.org/abs/1709.01749}{{\tt 1709.01749}}].

\bibitem{Ooguri:2016pdq}
H.~Ooguri and C.~Vafa,  {\em {Non-supersymmetric AdS and the Swampland}},
\href{http://www.arXiv.org/abs/1610.01533}{{\tt 1610.01533}}.

\bibitem{Louis:2015mka}
J.~Louis and S.~L{\"{u}}st,  {\em {Supersymmetric AdS\_7 backgrounds in
  half-maximal supergravity and marginal operators of (1,0) SCFTs}},
\href{http://www.arXiv.org/abs/1506.08040}{{\tt 1506.08040}}.

\bibitem{Cordova:2016xhm}
C.~Cordova, T.~T. Dumitrescu and K.~Intriligator,  {\em {Deformations of
  Superconformal Theories}}, JHEP {\bf 11} (2016) 135
[\href{http://www.arXiv.org/abs/1602.01217}{{\tt 1602.01217}}].

\bibitem{Freivogel:2016qwc}
B.~Freivogel and M.~Kleban,  {\em {Vacua Morghulis}},
\href{http://www.arXiv.org/abs/1610.04564}{{\tt 1610.04564}}.

\bibitem{Danielsson:2016mtx}
U.~Danielsson and G.~Dibitetto,  {\em {The fate of stringy AdS vacua and the
  WGC}},
\href{http://www.arXiv.org/abs/1611.01395}{{\tt 1611.01395}}.

\end{thebibliography}\endgroup

\end{document}